\documentclass[reprint,showpacs,amsmath,amssymb,pra]{revtex4-1}
\usepackage{graphicx}
\usepackage{dcolumn}
\usepackage{bm}

\begin{document}

\title{Optical kinks and kink-kink and kink-pulse interactions in resonant two-level media}

\author{Denis V. Novitsky}
\email{dvnovitsky@gmail.com} \affiliation{B. I. Stepanov Institute
of Physics, National Academy of Sciences of Belarus, Nezavisimosti
Avenue 68, BY-220072 Minsk, Belarus}

\date{\today}

\begin{abstract}
An optical kink is a shock-wave-like field structure which can
appear in a resonant two-level medium as a result of the nonlinear
process of self-steepening. We numerically simulate this process
using an adiabatically switching waveform as an input and confirm
the self-similarity of resulting kinks. The analysis is also
applicable to a more general waveform with a decaying trailing edge
which we call a kinklike pulse. We study in detail collisions of
kinks with other kinks and ultrashort pulses and demonstrate the
possibility to control kink speed by changing the parameters of
counterpropagating fields. The effects considered can be treated as
belonging to a wide class of unexplored phenomena in the regime of
incoherent light-matter interaction.
\end{abstract}


\maketitle

\section{Introduction}

An optical kink is a type of soliton which can be represented as a
shock wave (or shock front) with the shape preserved when
propagating in a nonlinear medium \cite{KivsharBook}. In the spatial
domain, such solitons are sometimes considered to be domain walls
separating distinct regions of space. Kinks were mathematically
introduced as solutions of the sine-Gordon equation \cite{Debnath}
and appear in different fields of physics. In the broad sense of the
word, kinks may represent not only ``true'' solitons with
particlelike properties but also solitonic waves which can interact
inelastically. In the context of optics, kinks were predicted and
observed in a number of nonlinear media. Optical shock solutions
were deduced for nonlinear interaction of waves via stimulated Raman
scattering in both nondispersive \cite{Christodoulides} and
dispersive media \cite{Agrawal} as well as in optical fibers
\cite{Kivshar93}. Kink solitons were also predicted to exist as
surface waves supported by an optical lattice imprinted in a
nonlinear medium \cite{Kartashov} and in a gain medium in the
presence of two-photon absorption \cite{Goyal}. As experimental
examples, we can mention pairs of kinks and antikinks obtained in
nonlinear photorefractive crystals \cite{Freedman}, dispersive shock
waves in optical fiber arrays \cite{Fatome}, and dark solitons
formed by domain walls in erbium-doped fiber lasers with
birefringent cavities \cite{Zhang1, Zhang2, Zhang3}.

In this paper, we are interested in another type of optical-kink
solution predicted by Ponomarenko and Haghgoo \cite{Ponomarenko}.
They have found such a solution considering the Maxwell-Bloch
equations which govern the propagation of light in resonant
two-level media. The kinks form as a result of self-steepening of
the input waveform, which should have a constant intensity at the
trailing edge. The characteristic time of the intensity jump of such
an optical shock is determined by the transverse relaxation time
$T_2$, i.e., by relaxation of microscopic polarization. On the other
hand, the longitudinal relaxation time $T_1$ governing the decay of
the excited-state population is responsible for kink disintegration.
It turns out that these solutions possess an important property of
self-similarity, i.e., scaling of their profiles during propagation.
Later, the same authors confirmed the preservation of this novel
type of kink under inhomogeneous broadening of the medium
\cite{Haghgoo}. It is worth noting that kink solutions are absent in
the media with the usual cubic (Kerr) nonlinearity and need more
sophisticated situations with competing nonlinear contributions,
such as cubic-quintic or saturable nonlinearity \cite{KivsharBook}.
The two-level medium, being an example of a saturable absorber,
supports the saturable nonlinearity which is responsible for kink
formation.

This paper can be considered a continuation of the work by
Ponomarenko and Haghgoo \cite{Ponomarenko} and, simultaneously, a
development of our previous studies of ultrashort pulses and their
interactions in two-level media \cite{Novit2011}. In particular,
unusual dynamics of matter was reported under asymmetric collisions
of self-induced transparency solitons \cite{Novit2012a} and in the
cases of extremely short (subcycle) pulses \cite{Novit2012b} and
chirped pulses \cite{Novit2016}. However, contrary to self-induced
transparency and other coherent effects which occur for ultrashort
pulses, kinks form in the regime of incoherent light-matter
interaction when, as stated above, the characteristic time of the
input field cannot be considered negligible in comparison to
relaxation times. On the other hand, consideration should differ
significantly from standard steady-state analysis, which in recent
decades allowed us to discover and study a number of effects such as
mirrorless optical bistability \cite{Hopf, Afan98} and local-field
effects \cite{Bowd93, Afan99}. Although we eventually deal with
stationary (constant-wave) fields, kinks are fundamentally dynamical
features. Moreover, kink formation can be treated as an example of a
transient process which needs full-scale modeling of temporal
dynamics. Transient processes of different natures have attracted
much attention in recent years, which ensures the relevance of our
study for modern nonlinear optics. From the viewpoint of solitonic
studies, the kinks can be considered another class of incoherent
solitons along with that reported in Ref. \cite{Afan02}.

Thus, in this paper, we numerically study the formation,
propagation, and interaction of optical kinks in a homogeneously
broadened two-level medium. The paper's structure is as follows. In
Sec. \ref{eqpars}, the main equations are given, and the parameters
of the calculations are discussed. Section \ref{form} is devoted to
the basic features of kink formation out of an input waveform; the
self-similarity is tested, and kinklike pulses are introduced. In
Secs. \ref{kki} and \ref{ksi}, collisions of kinks are studied with
counterpropagating kinks and ultrashort pulses (solitons),
respectively. The paper is completed with a short conclusion.

\section{\label{eqpars}Main equations and parameters}

In semiclassical approximation, light propagation in a homogeneously
broadened two-level medium can be described by the Maxwell--Bloch
equations. Since we deal with the incoherent regime of light-matter
interaction (i.e., the characteristic time of light-intensity change
is not much less than the medium relaxation times), we can safely
write these equations under the rotating-wave approximation as
follows (see, e.g., \cite{Novit2011}):
\begin{eqnarray}
\frac{dR}{d\tau}&=& i \Omega W + i R \delta - \gamma_2 R, \label{dPdtau} \\
\frac{dW}{d\tau}&=&2 i (\Omega^* R - R^* \Omega) - \gamma_1 (W-1),
\label{dNdtau} \\
\frac{\partial^2 \Omega}{\partial \xi^2}&-& \frac{\partial^2
\Omega}{\partial \tau^2}+2 i \frac{\partial \Omega}{\partial \xi}+2
i \frac{\partial \Omega}{\partial
\tau} \nonumber \\
&&=3 \epsilon \left(\frac{\partial^2 R}{\partial \tau^2}-2 i
\frac{\partial R}{\partial \tau}-R\right), \label{Maxdl}
\end{eqnarray}
where $\tau=\omega t$ and $\xi=kz$ are the dimensionless time and
distance; $\Omega=(\mu/\hbar \omega) E$ is the dimensionless
electric-field amplitude (normalized Rabi frequency); $E$ and $R$
are the complex amplitudes of the electric field and atomic
polarization, respectively; $W$ is the difference between
populations of ground and excited states; $\delta=\Delta
\omega/\omega=(\omega_0-\omega)/\omega$ is the normalized frequency
detuning; $\omega_0$ is the frequency of the atomic resonance;
$\omega$ is the light's carrier frequency; $\mu$ is the dipole
moment of the quantum transition; $\gamma_{1,2}=1/(\omega T_{1,2})$
are the normalized relaxation rates of population and polarization,
respectively; $\epsilon= \omega_L / \omega = 4 \pi \mu^2 C/3 \hbar
\omega$ is the dimensionless parameter of interaction between light
and matter (normalized Lorentz frequency); $C$ is the concentration
(density) of two-level atoms; $k=\omega/c$ is the wavenumber; $c$ is
the speed of light; and $\hbar$ is the Planck constant. An asterisk
stands for complex conjugation.

We solve Eqs. (\ref{dPdtau})--(\ref{Maxdl}) numerically using
essentially the same approach as in our previous publications (see
\cite{Novit2009}). The parameters of the medium and light used for
calculations (if not stated otherwise) are listed below. The
relaxation times $T_1=1$ ns and $T_2=0.1$ ps correspond to
semiconductors doped with quantum dots (although we do not consider
here the effects of the host medium, taking the background
dielectric permittivity to be unity), the detuning $\delta=0$ (exact
resonance, i.e., $\omega=\omega_0$), the central light wavelength
$\lambda=2 \pi c / \omega_0=0.8$ $\mu$m, and the strength of
light-matter coupling (the Lorentz frequency) $\omega_L=10^{11}$
s$^{-1}$. The region used in calculations includes a two-level
medium of thickness $L$ surrounded by vacuum regions of length $20
\lambda$ from both sides. The medium is supposed to be initially in
the ground state, so that $W(t=0)=1$.

\section{\label{form}Kinks and kink-like pulses}

\begin{figure}[t!]
\includegraphics[scale=0.85, clip=]{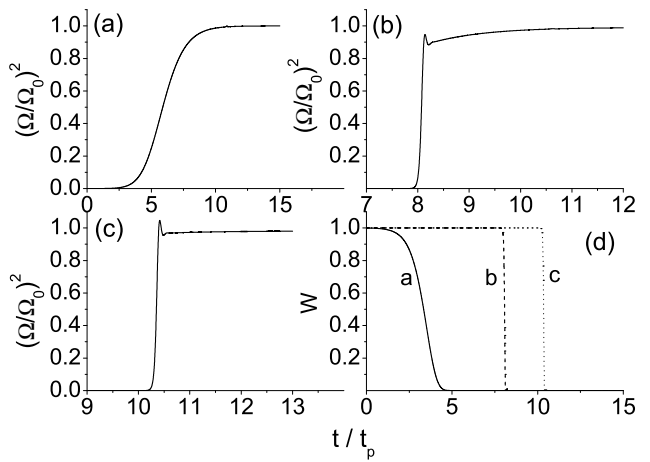}
\caption{\label{fig1} Light-intensity profiles at different
positions inside the medium: (a) $L=0$, (b) $L=500\lambda$, (c)
$L=1000\lambda$. (d) Dynamics of population difference at the same
positions as in (a)-(c). The parameters of input radiation are
$\Omega_0=0.5 \gamma_2$ and $t_p=50 T_2$.}
\end{figure}

In this section, we consider the process of kink formation as a
result of adiabatic switching of a constant wave (cw) which can be
described with a functional form as follows \cite{Ponomarenko}:
\begin{eqnarray}
\Omega(t)=\frac{\Omega_0}{1+e^{-(t-t_0)/t_p}}, \label{entform}
\end{eqnarray}
where $\Omega_0$ is the normalized amplitude of the cw field at the
trailing edge (Rabi frequency jump), $t_p$ is a switching time, and
$t_0$ is the offset time corresponding to the instant when the field
amplitude is half the maximum (further, we start calculations from
$t=0$ and set $t_0=5 t_p$ in this section). It is important that
this waveform has a constant amplitude at the trailing edge. To
obtain a characteristic example of the kink, we take the parameters
$\Omega_0=0.5 \gamma_2$ and $t_p=50 T_2$. Figure \ref{fig1} shows
the results of calculations of light-intensity dynamics at different
positions inside the medium: at $L=0$ (the entrance),
$L=500\lambda$, and $L=1000\lambda$ (the exit). It is seen that the
incident intensity changes smoothly according to the function
(\ref{entform}). As light propagates deep inside the medium, the
rising edge of the wave becomes more abrupt. This self-steepening of
the wave front is a characteristic feature of kink formation.
Analogous self-steepening occurs also in the dynamics of population
difference at corresponding depths inside the medium [Fig.
\ref{fig1}(d)]: the cw trailing edge rapidly saturates the medium,
so that the populations of both levels become equal (hence, $W=0$).
Our kinks have another peculiarity, an overshoot (or spike) at the
rising edge of the wave: for a very short time, the intensity
exceeds the steady-state level at the trailing edge. This feature
was absent in previous studies of kinks \cite{Ponomarenko} and,
perhaps, is due to chirping leading to modulations of the kink
profile. We discuss this point in more detail further.

\begin{figure}[t!]
\includegraphics[scale=0.85, clip=]{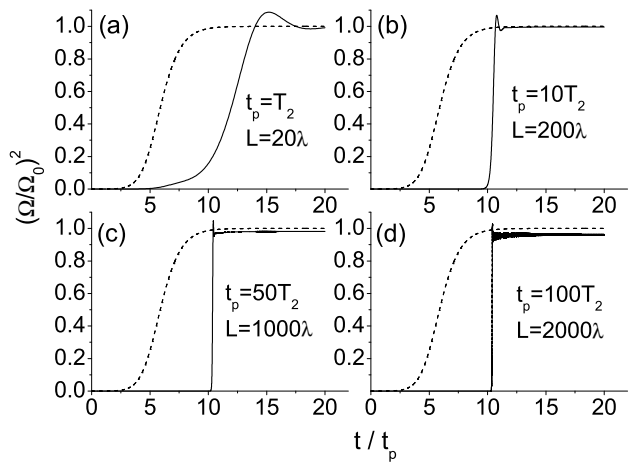}
\caption{\label{fig2} Light-intensity profiles for different
switching times $t_p$ and medium thicknesses $L$. The Rabi frequency
jump is $\Omega_0=0.5 \gamma_2$. Dashed line shows the input
waveform.}
\end{figure}

To verify that our waveforms are \textit{the} kinks, we should
demonstrate that they possess the property of self-similarity.
Rather than tracing the kink profile over large distances, let us
show that our waveforms can be scaled so that similar profiles can
be obtained at different properly chosen parameters. According to
Ref. \cite{Ponomarenko}, the distance of kink formation is given by
\begin{eqnarray}
L_*=\frac{\Omega_\infty^2 T_2^2}{\alpha}, \label{kinkdist}
\end{eqnarray}
where $\alpha$ is a linear absorption coefficient, and
$\Omega_\infty$ is the asymptotic kink amplitude (Rabi frequency
jump of the kink). Since $\Omega_\infty \sim \Omega_0$, $\alpha \sim
T_2$, and $t_p \sim T_2$, we can expect the self-similar resulting
waveforms after passing the distance
\begin{eqnarray}
L \sim \Omega_0^2 T_2 \sim \Omega_0^2 t_p. \label{simdist}
\end{eqnarray}
First, we fix the amplitude $\Omega_0$, so the input waves with
$t_{p1}$ and $t_{p2}$ should give the same kink profile after
propagating the distances related by $L_1/L_2=t_{p1}/t_{p2}$. Figure
\ref{fig2} verifies this expectation at $\Omega_0=0.5 \gamma_2$: the
kinks with $t_p=10 T_2$, $50 T_2$, and $100 T_2$ are very similar
and appear at the exit almost simultaneously (at $t=10 t_p$) after
propagating $L=200 \lambda$, $1000 \lambda$ and $2000 \lambda$,
respectively. This observation is equivalent to self-similarity
while changing the transverse relaxation rate $\gamma_2$ (or
corresponding time $T_2$). It is also seen that for a very rapid
change in the input intensity, self-steepening is absent [Fig.
\ref{fig2}(d)] since the switching time is already of the same order
of magnitude as the temporal width of the kink ($t_p \sim T_2$).
According to the scaling law, self-steepening could develop already
after $L=20 \lambda$ but is absent even after a tenfold increase in
the distance. This is direct corroboration of the expected condition
$t_p \gg T_2$ for kink formation. Note also that the overshoot at
the rising edge of the waveform is present in this case as well.

\begin{figure}[t!]
\includegraphics[scale=0.85, clip=]{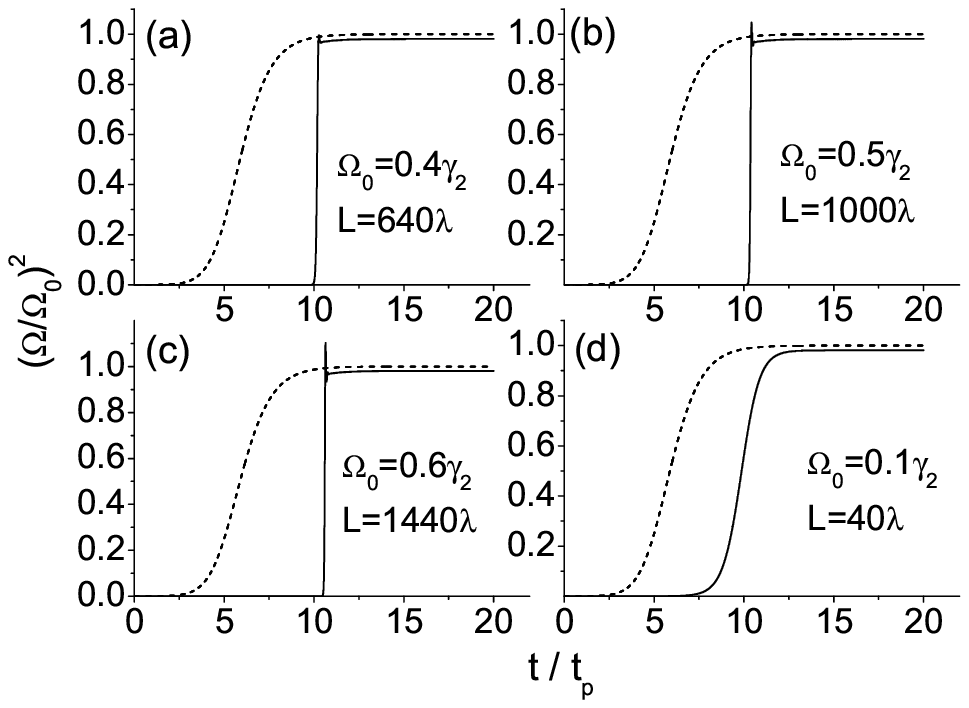}
\includegraphics[scale=0.95, clip=]{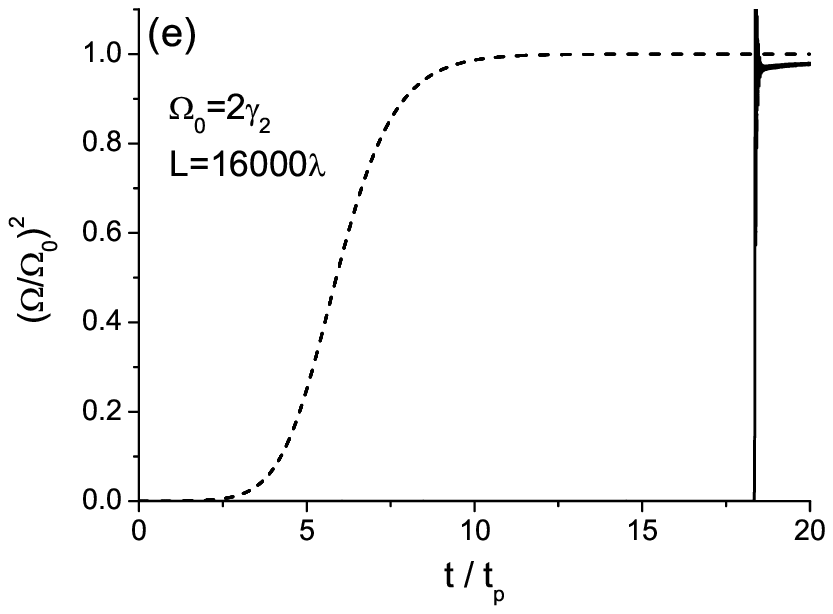}
\caption{\label{fig3} Light-intensity profiles for different Rabi
frequency jumps $\Omega_0$ and medium thicknesses $L$. The switching
time is $t_p=50 T_2$. The dashed line shows the input wave-form.}
\end{figure}

Second, we fix the switching time ($t_p=50 T_2$), and according to
Eq. (\ref{simdist}), the input waves with $\Omega_{0,1}$ and
$\Omega_{0,2}$ should give the same kink profile after passing the
distances related by $L_1/L_2=(\Omega_{0,1}/\Omega_{0,2})^2$. The
results of testing this expectation are given in Fig. \ref{fig3},
where the kinks with $\Omega_0=0.4 \gamma_2$, $0.5 \gamma_2$, and
$0.6 \gamma_2$ are shown after propagating $L=640 \lambda$, $1000
\lambda$, and $1440 \lambda$, respectively. One can see that these
waveforms are very similar and need almost the same time to pass the
medium (some discrepancy can be attributed to the fact that the
relation $\Omega_\infty \sim \Omega_0$ is not exact). We should also
note that the waveforms with higher intensity move much faster than
less powerful ones. Similar kinks form at $\Omega_0=0.3 \gamma_2$
($L=360 \lambda$) and $\Omega_0=0.2 \gamma_2$ ($L=160 \lambda$), but
as shown in Fig. \ref{fig3}(d), the input waveform with
$\Omega_0=0.1 \gamma_2$ cannot produce a kink after the distance
$L=40 \lambda$ (or at longer distances). This is in accordance with
the prediction of the critical amplitude, which gives the lower
bound for kink existence \cite{Ponomarenko}.

\begin{figure}[t!]
\includegraphics[scale=0.95, clip=]{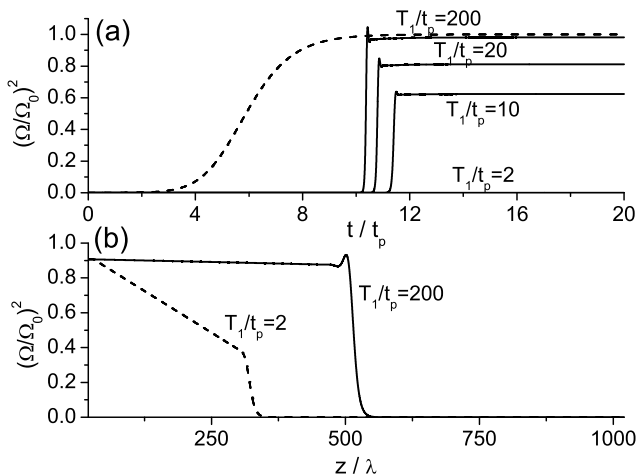}
\caption{\label{fig4} (a) Light-intensity profiles and (b) intensity
distributions along the medium (at time point $t=10 t_p$) for
different longitudinal relaxation times $T_1$. Other parameters are
$\Omega_0=0.5 \gamma_2$, $t_p=50 T_2$, $L=1000\lambda$. The dashed
line shows the input waveform.}
\end{figure}

On the other hand, the upper bound is given by the expression
$\Omega_0 \leq 0.5 \gamma_2$ \cite{Ponomarenko}. At larger
amplitudes, according to Ref. \cite{Ponomarenko}, the kink should
become a chirped wave due to the onset of Rabi oscillations at
larger Rabi frequencies. However, we do not see any fundamental
changes in its profile at $\Omega_0=0.6 \gamma_2$ [Fig.
\ref{fig3}(c)]. Perhaps, the chirp appears as the spike at the
rising edge of the kink which already exists at $\Omega_0=0.5
\gamma_2$ and gradually diminishes with decreasing $\Omega_0$.
According to our calculations, this spike appears long before the
bound value of Ref. \cite{Ponomarenko}: for example, it is clearly
seen at $\Omega_0=0.4 \gamma_2$ as well [Fig. \ref{fig3}(a)]. Thus,
our numerical results show that the Rabi-oscillation-induced chirp
does not become apparent abruptly at $\Omega_0=0.5 \gamma_2$, but
gradually becomes more pronounced.

What seems to be more important is that the kinks at larger
$\Omega_0$ lose the property of self-similarity despite remaining
stable waveforms. This conclusion is illustrated in Fig.
\ref{fig3}(e), which shows kink formation at $\Omega_0=2 \gamma_2$
after propagating the distance $L=16000\lambda$. According to Eq.
(\ref{simdist}), this waveform should be similar to those given in
Figs. \ref{fig3}(a)-\ref{fig3}(c), but it is not. The former need
the same time ($\sim 10t_p$) to pass the medium, while the latter is
much slower than expected (it needs about $17t_p$). Note that some
indication of this effect was present already in Fig. \ref{fig3}(c),
where slightly more time was required for the kink with
$\Omega_0=0.6 \gamma_2$ to propagate through the medium. Therefore,
the value $\Omega_0=0.5 \gamma_2$ can indeed be viewed as an upper
bound for \textit{self-similar} kinks.

The critical lower bound of the amplitude is a function of the
longitudinal relaxation time $T_1$ \cite{Ponomarenko}. However, we
will not study that dependence here. Rather, let us see how $T_1$
influences the kink appearance at the fixed amplitude, switching
time, and medium thickness. The results of calculations at different
values of $T_1$ are shown in Fig. \ref{fig4}. As the longitudinal
relaxation time gets lower, the resulting Rabi frequency jump of the
kink decreases as well as its speed. Finally, when $T_1 \sim t_p$,
the kink entirely disappears due to energy dissipation. The decay of
the field inside the medium in this case is illustrated in Fig.
\ref{fig4}(b), in contrast to the propagating kink in the case $T_1
\gg t_p$. The dissipation of the kink can be equivalently obtained
when we leave the relaxation times unchanged and consider very
slowly switching waves with $t_p$ of the same order of magnitude as
$T_1$.

\begin{figure}[t!]
\includegraphics[scale=1, clip=]{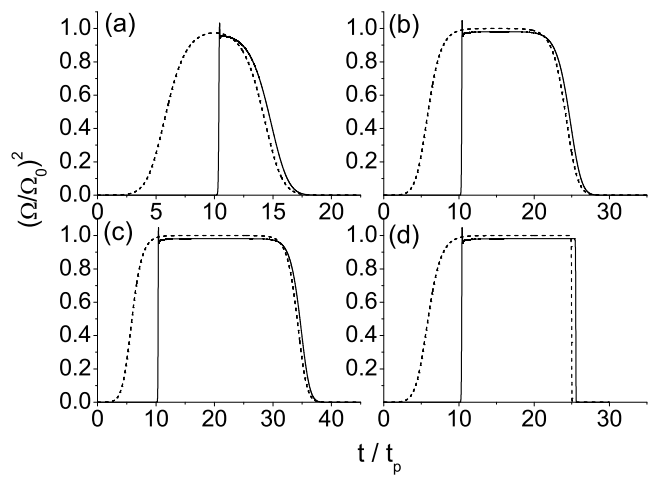}
\caption{\label{fig5} Light-intensity profiles of the kink-like
pulses. The parameters used are $\Omega_0=0.5 \gamma_2$,
$L=1000\lambda$, $t_p=50 T_2$, and (a) $t'_p=50 T_2$, $t'_0=15 t_p$;
(b) $t'_p=50 T_2$, $t'_0=25 t_p$; (c) $t'_p=50 T_2$, $t'_0=35 t_p$;
and (d) $t'_p=T_2$, $t'_0=25 t_p$. Dashed lines show the input
waveforms.}
\end{figure}

\begin{figure*}[t!]
\includegraphics[scale=1, clip=]{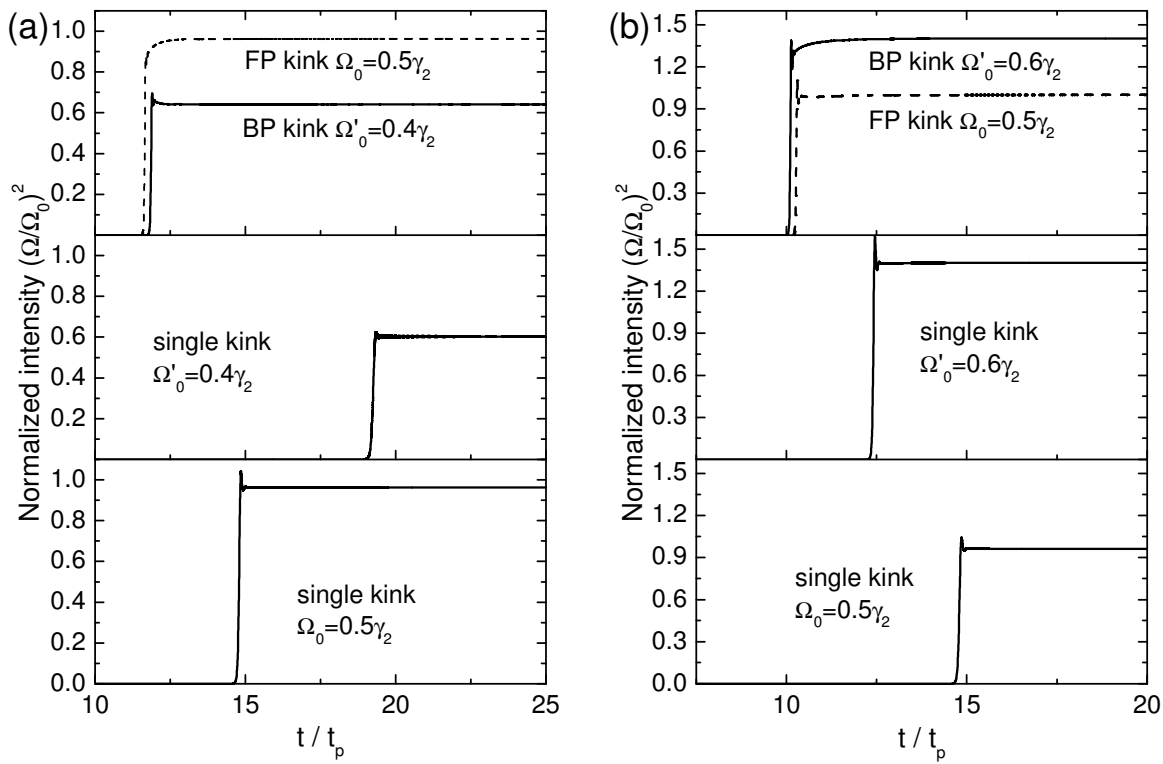}
\centering \caption{\label{fig6} Intensity profiles in the case of
an interaction between the forward-propagating (FP) kink of
amplitude $\Omega_0=0.5 \gamma_2$ and the backward-propagating (BP)
kink of amplitude (a) $\Omega'_0=0.4 \gamma_2$ and (b)
$\Omega'_0=0.6 \gamma_2$ in comparison with the single-kink results.
Other parameters are medium thickness $L=2000\lambda$ and switching
time $t_p=t'_p=50 T_2$ for both kinks.}
\end{figure*}

Although it was stated that the constant intensity at the trailing
edge of the input wave is a necessary condition for kink formation,
this requirement is not absolute for self-steepening development. In
fact, the input wave can be switched off after some time. It turns
out that this time can be rather short. We model switching on and
off the input light with the waveform as follows:
\begin{eqnarray}
\Omega(t)=\frac{\Omega_0}{(1+e^{-(t-t_0)/t_p})(1+e^{(t-t'_0)/t'_p})},
\label{entform1}
\end{eqnarray}
where $t'_p$ and $t'_0$ are the switching-off time and the
corresponding offset time, respectively, which generally differ from
the switching-on characteristics $t_p$ and $t_0$. Figures
\ref{fig5}(a)-\ref{fig5}(c) show that the shock is formed at the
rising edge of the waveform (\ref{entform1}) even for rather short
switching-off offset $t'_0=15 t_p$ (recall that the switching-on
value is $t_0=5 t_p$) and remains essentially the same at $t'_0=25
t_p$ and $t'_0=35 t_p$. One can see that decay of the transmitted
radiation perfectly replicates the switching-off dynamics of the
input waveform, even for very fast switching off [$t'_p=T_2$, Fig.
\ref{fig5}(d)]. We call such profiles formed as a result of
transmission of the waveforms (\ref{entform1}) \textit{the kinklike
pulses}.

\section{\label{kki}Kink-kink interactions}

In this section, we discuss interactions between kinks or kinklike
pulses. One can imagine two situations: copropagating and
counterpropagating kinks. In the case of copropagating waveforms
launched one after another in the same direction, the perspectives
are very limited. The first waveform transforms into a kink and
saturates the medium, so that the levels are equally populated, and
hence, $W=0$. Then, the second waveform propagates unchanged: since
the medium is saturated, self-steepening does not occur, and the
kink does not form. If the first waveform is a kinklike pulse
(\ref{entform1}), after its passing, the medium slowly returns to
the ground state (with the characteristic time $T_1$). This means
that the second waveform can be transformed into a kink if it is
launched in a period long enough after the first kinklike pulse.

Let us consider a more interesting situation in which the waveforms
propagate through the two-level medium in opposite directions and
meet inside it. Examples of such a collision are given in Fig.
\ref{fig6}, where one of the input fields has the Rabi frequency
jump $\Omega_0=0.5 \gamma_2$, while the counterpropagating field is
either less ($\Omega'_0=0.4 \gamma_2$) or more ($\Omega'_0=0.6
\gamma_2$) powerful than the first one. We choose the thickness to
be large enough ($L=2000\lambda$) that the kinks have enough time to
form before collision. We compare the case of collision (top panels)
with the case of single-kink formation and propagation through the
medium (middle and bottom panels). One can see that the collision
practically does not change the intensity of both kinks. However,
the propagation speed of the kinks dramatically increases. Moreover,
both kinks now move almost at the same speed. For example, the
single wave-forms with $\Omega_0=0.5 \gamma_2$ and $\Omega'_0=0.4
\gamma_2$ need times of approximately $15t_p$ and $19t_p$ to pass
the medium, while they need only about $12t_p$ in the case of the
collision. Thus, we have two important facts: (i) an increase in
speed propagation, and (ii) equalization of the speeds of both
kinks.

\begin{figure}[t!]
\includegraphics[scale=1, clip=]{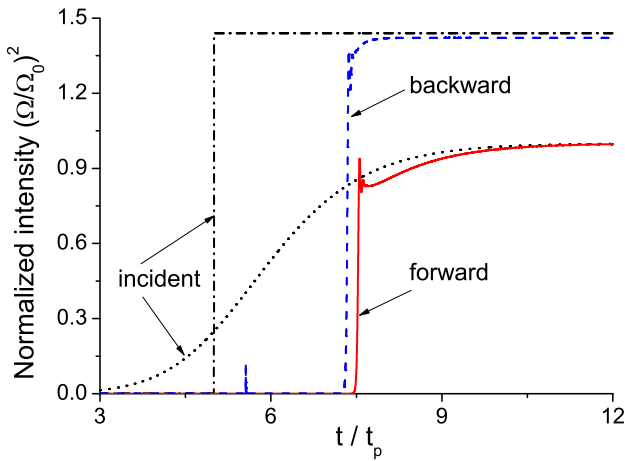}
\caption{\label{fig7} (Color online) Intensity profiles in the case
of an interaction between the forward-propagating (FP) kink of
amplitude $\Omega_0=0.5 \gamma_2$ and the backward-propagating (BP)
wave front of amplitude $\Omega'_0=0.6 \gamma_2$. Other parameters
are medium thickness $L=1000\lambda$ and switching time $t_p=50 T_2$
for the FP waveform and $t'_p=0.001 T_2$ for the BP waveform.}
\end{figure}

\begin{figure}[t!]
\includegraphics[scale=0.9, clip=]{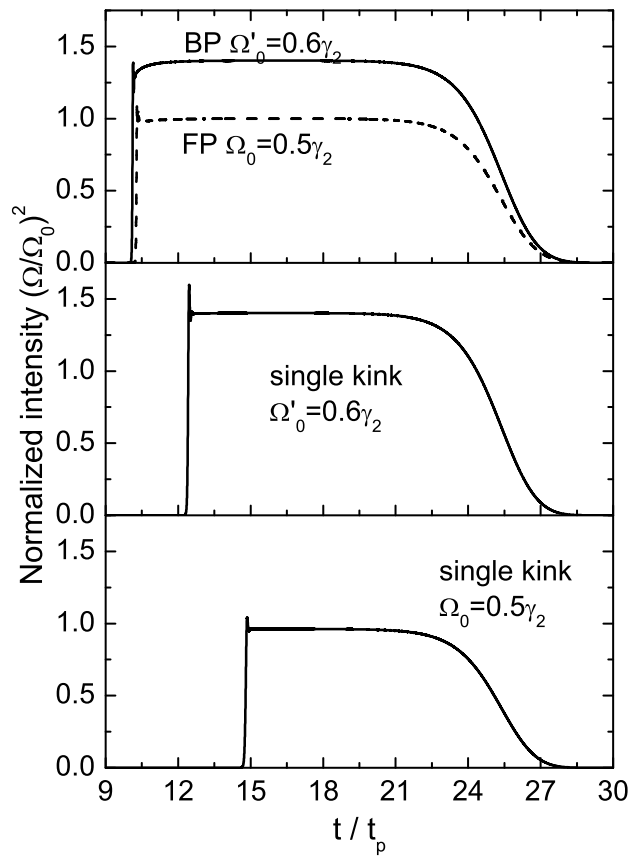}
\caption{\label{fig8} Intensity profiles in the case of an
interaction between the forward-propagating (FP) kinklike pulse of
amplitude $\Omega_0=0.5 \gamma_2$ and the backward-propagating (BP)
kinklike pulse of amplitude $\Omega'_0=0.6 \gamma_2$. Other
parameters are medium thickness $L=2000\lambda$ and switching time
$t_p=t'_p=50 T_2$ for both kinklike pulses.}
\end{figure}

\begin{figure}[t!]
\includegraphics[scale=0.95, clip=]{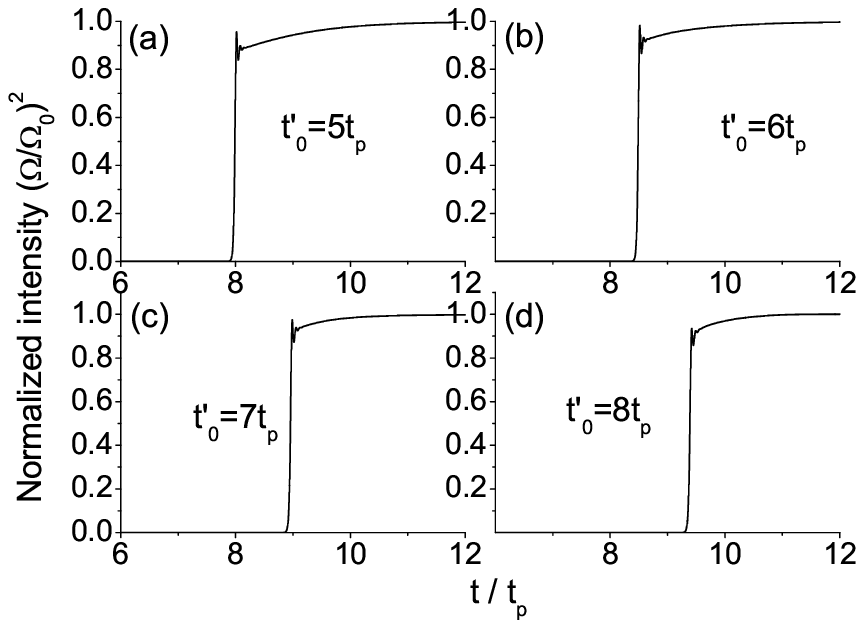}
\caption{\label{fig9} Intensity profiles of the forward-propagating
kink (amplitude $\Omega_0=0.5 \gamma_2$, offset time $t_0=5 t_p$)
after an interaction with the backward-propagating kink of amplitude
$\Omega'_0=0.6 \gamma_2$ and different offset times $t'_0$. Other
parameters are medium thickness $L=1000\lambda$ and switching time
$t_p=t'_p=50 T_2$ for both kinks.}
\end{figure}

This observation is due to several reasons. First, the speed of the
kink depends on the stationary field intensity (Rabi frequency
jump). After the collision, the kink propagates through the spatial
region where the background stationary field is created by the
counterpropagating waveform. In other words, the resulting intensity
of the field after collision is the same and is given by the sum of
amplitudes of the kinks. The raised sum intensity results in an
increase in the propagation speed of both kinks. Second, the
interaction of counterpropagating waves leads to their interference
and hence to the creation of periodic grating of the population
difference inside the medium \cite{Narducci} and even to effects of
wave instability \cite{Bar-Joseph}. The equal speeds of both kinks
can be attributed to strong interaction via this population grating,
which effectively equalizes the speed of the waveforms.

Similar observations are valid for other situations, e.g., for two
counterpropagating wave fronts of abruptly (not adiabatically)
switched cw fields. In a sense, the increased speed of wave fronts
moving apart after a collision can be interpreted as a repulsion of
wave fronts or kinks. This transient process deserves a separate
detailed study. Here, suffice it to say that we obtained the
instrument to control the speed of the kinks which has the most
impressive manifestations of kinks of comparatively low intensity:
the increase in their speed after collision with more powerful kinks
is the most striking.

Moreover, it is not necessary to use the kink as a
backward-propagating waveform to control the forward-propagating
kink. Other variants are possible as well. In Fig. \ref{fig7}, the
results for the collision of the kink with the counterpropagating
wave front switching almost instantaneously ($t'_p=0.001 T_2$) are
depicted. It is seen that both waveforms pass a medium of thickness
$L=1000 \lambda$ almost simultaneously and need only about $7.5 t_p$
[compare with the more than $10 t_p$ needed for a single kink to
propagate through a medium of the same length; see Fig.
\ref{fig3}(b)]. Another possibility is to use not the kinks, but
kinklike pulses. The situation shown in Fig. \ref{fig8} perfectly
corresponds to the case in Fig. \ref{fig6}(b) where the kinks were
considered. There is no need to say that we could analyze other
analogous combinations, such as ``a kink + a kinklike pulse'' or ``a
kinklike pulse + a cw front''. All these situations have common
features: interference of counterpropagating waves and formation of
population grating.

Thus, kink propagation can be controlled with the counterpropagating
waveform. What instruments do we possess to change the speed of the
kink? The first such instrument is the intensity of the control
(counterpropagating) waveform. This is illustrated in Fig.
\ref{fig6}: the kink with the Rabi frequency jump $\Omega_0=0.5
\gamma_2$ passes the medium faster after interaction with a more
powerful kink ($\Omega'_0=0.6 \gamma_2$) than with a less powerful
one ($\Omega'_0=0.4 \gamma_2$). Another instrument is the offset
time $t'_0$ of the counterpropagating waveform. As an example, Fig.
\ref{fig9} shows that increasing this offset time from $5 t_p$ (the
backward-propagating kink is launched simultaneously with the
forward-propagating one) to $8 t_p$ (the backward-propagating kink
is launched later than the forward-propagating one by $3 t_p$)
results in growing the passage time of the forward-propagating kink
from about $8 t_p$ to almost $9.5 t_p$. The reason is obvious: the
waveforms collide later, so the kink mostly propagates alone, and
its passage time tends to that of a single kink (slightly over $10
t_p$).

\section{\label{ksi}Kink-pulse interaction}

In this section, we study the interaction of kinks with
counterpropagating ultrashort pulses. In particular, we consider
pulses of Gaussian profile $\Omega=\Omega_p \exp[-(t-t'_0)^2/2
t'^2_p]$ with a duration $t'_p$ and an offset time $t'_0$. It is
well known that, in the regime of coherent light-matter interaction
(when $t'_p \ll T_2 \ll T_1$), such pulses form self-induced
transparency (SIT) solitons \cite{McCall}. The pulses considered
here approximately correspond to such solitons (although the
duration is only one or two orders of magnitude less than the
relaxation time $T_2$). The key parameter of SIT solitons is their
area, which at the constant $t'_p$, can be treated as a measure of
pulse amplitude $\Omega_p$. If the area $A$ is equal to $2 \pi$,
such a pulse inverts the medium at the rising edge and then returns
it back exactly to the ground state at the trailing edge. The pulse
with area differing from $2 \pi$ leaves the medium partially excited
and, according to the ``area theorem'', can be transformed into a $2
\pi$-soliton as it propagates deep inside the medium.

\begin{figure}[t!]
\includegraphics[scale=1, clip=]{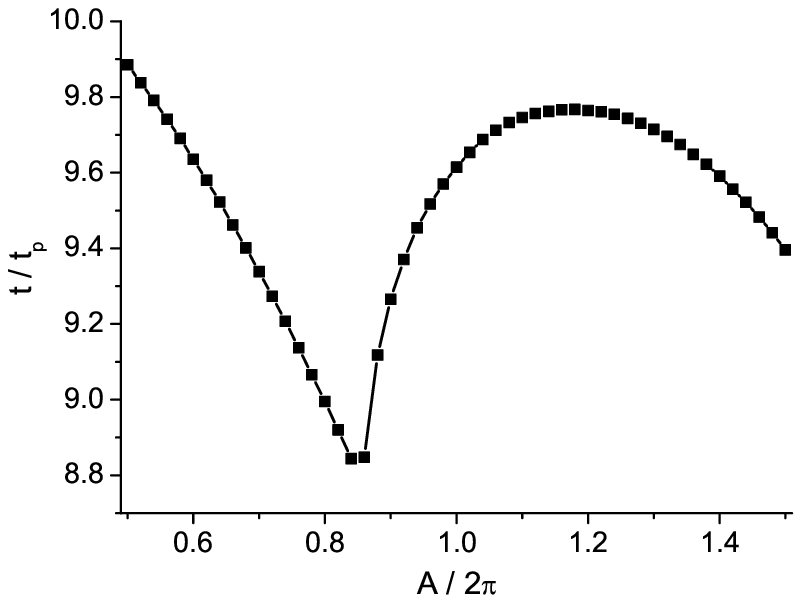}
\caption{\label{fig10} Time interval needed for the
forward-propagating kink to pass the medium as a function of the
area of the backward-propagating pulse. Medium thickness
$L=1000\lambda$, kink switching time $t_p=50 T_2$, and Rabi
frequency jump $\Omega_0=0.5 \gamma_2$; pulse duration is $0.1 T_2$,
and offset time $t'_0=t_p$.}
\end{figure}

\begin{figure}[t!]
\includegraphics[scale=0.85, clip=]{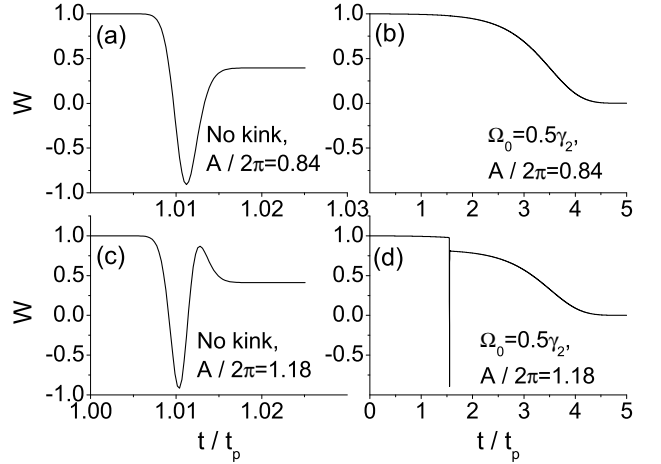}
\caption{\label{fig11} Dynamics of population difference for (a) and
(c) a single pulse and (b) and (d) kink-pulse interaction. Pulses of
two areas are considered: (a) and (b) $A/2\pi=0.84$, (c) and (d)
$A/2\pi=1.18$. The other parameters are the same as in Fig.
\ref{fig10}.}
\end{figure}

\begin{figure}[t!]
\includegraphics[scale=1, clip=]{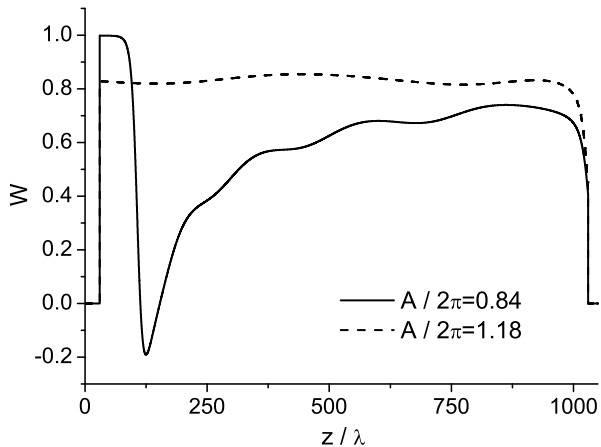}
\caption{\label{fig12} Distribution of population difference along
the medium for a single pulse at the time instant $t=4t_p$. Pulses
of two areas are considered: $A/2\pi=0.84$ and $1.18$. The other
parameters are the same as in Fig. \ref{fig10}. Note also the vacuum
regions of length $20 \lambda$ from both sides of the two-level
medium.}
\end{figure}

\begin{figure}[t!]
\includegraphics[scale=1, clip=]{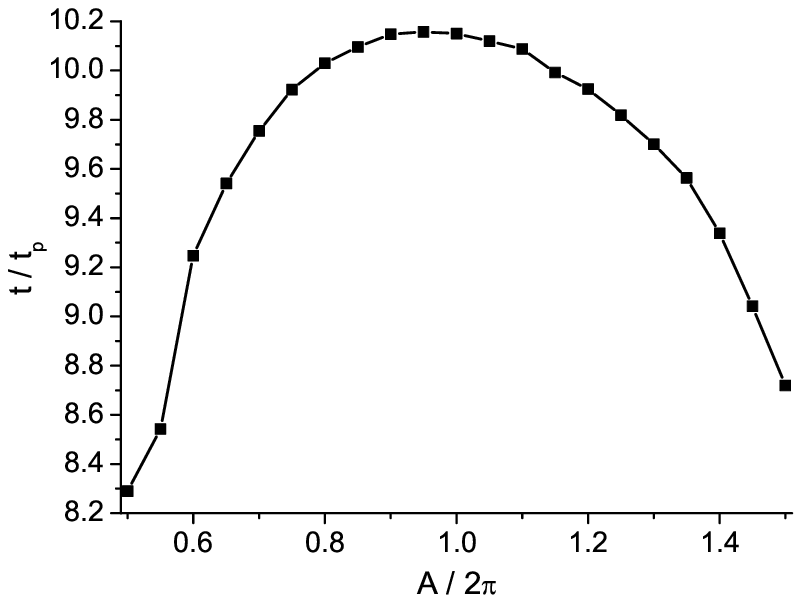}
\caption{\label{fig13} Time interval needed for the
forward-propagating kink to pass the medium as a function of the
area of the backward-propagating pulse. Medium thickness
$L=1000\lambda$, kink switching time $t_p=50 T_2$, and Rabi
frequency jump $\Omega_0=0.5 \gamma_2$; pulse duration is $0.01
T_2$, and offset time $t'_0=t_p$.}
\end{figure}

Let us consider the interaction of the waveform (\ref{entform}) with
counterpropagating pulses of different area with the duration
$t'_p=0.1 T_2$ and the same offset time $t'_0=t_p$. As in the
previous section, we focus on kink speed as a parameter influenced
by collision. Figure \ref{fig10} shows the dependence of kink
passage time on the pulse area. It is seen that there is a strong
minimum around pulse area $A/2 \pi=0.84$ when the kink needs
approximately $8.8 t_p$ to pass a medium of $L=1000 \lambda$. At the
maximum, around $A/2 \pi=1.18$, the passage time grows to $9.8 t_p$,
which with the parameters used, is almost equal to the value in the
case of a single kink. Thus, pulse area (hence, intensity) can be
considered an instrument to control the kink speed.

To get insight into the interaction process, we plot in Fig.
\ref{fig11} the population difference at the entrance of the medium
for a single pulse (no kink, left panels) and in the presence of a
kink (right panels). The cases of the minimum and maximum of the
curve in Fig. \ref{fig10} are considered. It is seen that single
pulses give approximately the same final level of population
difference (around $0.4$), although the one with area $A/2 \pi=1.18$
makes a complete cycle of excitation and deexcitation [Fig.
\ref{fig11}(c)], in contrast to the one with area $A/2 \pi=0.84$
[Fig. \ref{fig11}(a)]. This difference between the pulses turns out
to be the key factor governing their different interactions with the
kink. There is a sharp dip induced by the pulse with area $A/2
\pi=1.18$ and superposed on the gradual saturation of the population
difference due to the kink [Fig. \ref{fig11}(d)]. On the contrary,
the pulse with area $A/2 \pi=0.84$ does not have any visible
influence on the saturation of the medium caused by the kink [Fig.
\ref{fig11}(b)].

The reason is that the pulse with lower area has another fate even
before reaching the entrance of the medium. This suggestion is
confirmed by an analysis of medium excitation created by single
pulses and experienced by kinks. In Fig. \ref{fig12}, the population
difference along the medium is shown at the time instant $t=4t_p$
when the backward-propagating pulses should already be out of the
medium. The drastic difference in behavior of pulses with areas $A/2
\pi=1.18$ and $0.84$ is clearly seen. The first leaves the medium
almost uniformly excited at the level of $W \sim 0.8$. The second
one is strongly absorbed near the medium entrance, resulting in a
profound dip while the medium near the exit (where the pulse was
launched) is excited almost to the same low level as in the case of
$A/2 \pi=1.18$. This is due to the fact that the condition $t'_p=0.1
T_2$ does not provide a coherent regime of light-matter interaction
and formation of SIT solitons. A waveform giving rise to a kink
needs more time for self-steepening and medium saturation when
propagating through the strongly and nonuniformly excited medium
(the case of $A/2 \pi=0.84$) than in the case of a weakly and
uniformly excited one ($A/2 \pi=1.18$).

For pulses of shorter duration ($t'_p=0.01 T_2$) which better
correspond to the condition of a coherent regime, we have a more
obvious and predictable picture (Fig. \ref{fig13}). The $2 \pi$
pulses which leave the medium almost unperturbed and easily
transform into SIT solitons have minimal influence on the kink
speed, whereas propagation of the pulses with area of $\pi$ or $3
\pi$ results in the maximum level of medium excitation and gives the
strongest kink retardation (down to $8.3 t_p$). Thus, using
ultrashort counterpropagating pulses is another tool for controlling
kink passage through the medium with the possibility to change the
resulting kink speed with the proper choice of pulse duration and
area.

\section{\label{concl}Conclusion}

Using numerical simulations, we have studied in detail the formation
of optical kinks and kinklike pulses in two-level media and
demonstrated their self-similarity. The results are generally in
agreement with the analytical theory of Ponomarenko and Haghgoo
\cite{Ponomarenko}, although there are some minor discrepancies. In
particular, the chirped kink appears in our calculations in the
range of parameters where analytical theory predicts a monotonic
waveform. This difference can be ascribed to our more general
approach since we do not use the slowly-varying-envelope
approximation. Further, we have studied collisions of kinks with
other waveforms and shown that the speed of kink propagation can be
significantly changed by interaction with a counterpropagating kink,
stationary wave front, or ultrashort pulse. These possibilities are
interesting from the physical point of view but can also be used in
different optical schemes for nonlinear control of propagating wave
fields and optical information processing. In particular, kinks can
be considered as a substitute for the usual (pulse) solitons as
optical bits of information since they can exist in different
parameter ranges. We should also note that the inelasticity of
collisions resulting in a speed change does not allow us to consider
such kinks as strict solitons, only as solitonic waves.

In conclusion, we make several remarks on possible directions of
kink studies. Although it was reported that kinks are preserved
under inhomogeneous broadening \cite{Haghgoo}, it may be interesting
to test this prediction with our more general numerical approach
adapting the scheme described in Ref. \cite{Novit2014}. Since
calculations in this paper were performed for two-level atoms in
vacuum, a more general consideration is worth exploring, taking into
account the background matrix (see, e.g., the study of its influence
on mirrorless optical bistability \cite{Novit2011a}). In a dense
enough medium, near-dipole-dipole interactions between active
particles begin to play an essential role, which can be taken into
account through the so-called local-field correction
\cite{Novit2010}. All this allows us to consider optical kinks an
interesting example of incoherent phenomena lying between and
connecting the regime of ultrafast processes (with self-induced
transparency solitons being a characteristic feature) and the
stationary regime (with optical bistability and similar effects). Of
course, it is not less important to obtain these kinks
experimentally since there are only a few practical realizations of
optical shock waves and studies of their rich dynamics. The
solid-state systems are especially attractive for experimental
observations, in particular solids doped with resonant atoms (e.g.,
rare-earth ions) and bulk semiconductors doped with quantum dots.

\end{document}